\documentstyle[12pt]{article}
\newcounter{numerone}

\newcommand {\cJ}{{\cal J}}

\newcommand {\cL}{{\cal L}}
\newcommand {\cM}{{\cal M}}

\newcommand{\bR}{{\bf R}}

\def\b{\beta}

\def\d{\delta}

\def\l{\lambda}
\def\m{\mu}

\def\t{\tau}

\newcommand{\sect}[1]{\setcounter{equation}{0}\section{#1}}
\renewcommand{\theequation}{\thesection.\arabic{equation}}
\newcommand{\be}{\begin{equation}}
\newcommand{\ee}{\end{equation}}
\newcommand{\bea}{\begin{eqnarray}}
\newcommand{\eea}{\end{eqnarray}}

\begin{document}
\begin{titlepage}

\begin{flushright}
QFT-TSU-9/96 \\
hep-th/9607114\\
\end{flushright}

\begin{center}
\large{{\bf On the Minimal Model of Anyons}} \\
\vspace{1.0cm}

\large{ I.V. Gorbunov\footnote{E-mail: ivan@phys.tsu.tomsk.su},
S.M. Kuzenko and S.L. Lyakhovich
}\\

\footnotesize{{\it Department of Physics, Tomsk State University\\
Lenin Ave. 36, Tomsk 634050, Russia} } \\
\end{center}
\vspace{1.5cm}

\begin{abstract}
We present new geometric formulations for the fractional spin particle models
on the minimal phase spaces. New consistent couplings of the anyon to
background fields are constructed. The relationship between our approach and
previously developed anyon models is discussed.
\end{abstract}
\vspace{40mm}

\begin{flushleft}
July 1996
\end{flushleft}

\vfill
\null
\end{titlepage}

\section{Introduction} \label{s1}

Particles with fractional statistics, so called anyons
\cite{LeinaasMyrheim71Wilczek82,Wilczekbook90},
living in $(1+2)$--dimensional space--time have
attracted a considerable interest in recent years.
This interest basically ensues from the important role which the
anyons were established to play in some planar physics phenomena, like
the fractional quantum Hall effect
\cite{Laughlin83},
high--$T_c$ superconductivity
\cite{Wilczekbook90} as well as in some
$d=3$ quantum field models [4-6].

The most popular approach to the description of anyons is to employ the
Chern--Simons gauge field to generate statistics of the particles [6-9].
Up to now, however,
it is not quite clear whether the fractional
statistics states may emerge in this approach as the result of
interaction with the gauge field only or this is an inherent quality
of the particles themselves.  Even in the latter case, it
would not be possible to eliminate the Chern--Simons field without
violation of locality \cite{FroMar8891Ger91}.

Another traditional method to describe anyons, which we follow
in the present paper, is the group--theoretical approach [10-20].
This implies to derive the quantum theory by quantizing
a classical mechanics model based on some
Poincar\'e--invariant Lagrangian. Within this approach the possibility for
the particle spin to be fractional is made evident from the fact that
the universal covering map for the $(1+2)$--dimensional Poincar\'e\, group
is infinite-sheeted.

In constructing the mechanics models of spinning particles, the phase space
is usually extended by some internal variables destined to realize the spin
part of the Lorentz generators (see, e.g.,
\cite{MarneliusMartensson91} and references therein).
At that an appropriate set of constraints should be imposed on the phase
space variables in order to obtain the irreducible dynamics of spinning
particle with correct number of degrees of freedom. In particular, the
particle mass and spin are to be identically conserved on the constrained
surface.
Of especial interest are minimal models defined to contain a minimal
number of constraints providing the identical conservation of the
Casimir functions associated with the phase space Poincar\'e generators.
Such models are, in a sense, universal since any correct spinning particle
model should turn into a minimal one upon restricting the dynamics to a
surface of constraints. Another important peculiarity of the models
consists in strong restrictions the requirement of minimality imposes
on the topological structure of the phase space. In particular, its
dimension proves to be uniquely determined from a simple counting of
degrees of freedom.

For $d=3$ any minimal model is seen to be equivalent to a dynamical
system with two first-class constraints corresponding to the
Casimir operators
of the Poincar\'e group $ISO(1,2)$. Since for $d=3$ the spinning
particle possesses the same number of degrees of freedom as a spinless
one, the internal (spin) sector of the phase space has
to be two-dimensional.
A useful classification of $d=3$ minimal models has been recently given
\cite{CortesPlyushchay95hep}.
It is based on the observation that the phase space functions $M^aM_a$,
$M_a$ being the spin part of the Lorentz generators, commute with the
Poincar\'e generators, with respect to the Poisson bracket,
hence
\begin{equation}
M^aM_a =C=const             \label{1}
\end{equation}
is a constant parameter of the model. If $M_a$ are realized in the form
$M_a = M_a(z_i)$, where $z_i$, $i=1,2$, are internal phase space
variables, then Eq. (\ref{1}) implies that the topology of the spin space is
determined by the value of $C$. For $C<0$ we have a two-sheeted
hyperboloid, for $C>0$ a one-sheeted hyperboloid which turns into a cone
for $C=0$. It has been shown \cite{CortesPlyushchay95hep}
how to realize the minimal model for
each value of $C$ by describing the spin degrees of freedom in terms
of the constrained variables $M_a$ required to form the Lorentz algebra
with respect to the Poisson bracket.

In the present paper we show that all minimal
models uniquely originate from natural physical and
geometrical principles. The starting points
of our approach are as follows. First, the spin space is the cotangent
bundle $T^*(S^1)$ of a one-sphere for $C \geq 0$ and coincides with
Lobachevsky space $\cL$ for $C < 0$. Second, the action functional
should be specified in terms of geometric invariants related to the
phase space $T^*(\bR^{1,2} \times S^1)$ for $C \geq 0$ and
$T^*(\bR^{1,2}) \times \cL$ for $C < 0$. Our approach allows to obtain new
insights into the structure of the minimal models. In particular,
we construct new consistent couplings of the anyon to background field.
We also demonstrate canonical equivalence of some models previously
considered to describe different quantum dynamics.

There exist several general approaches to the quantization of
particles with fractional spin. In the framework of the
group--theoretical approach
[10,17-19]
one should use the infinite-dimensional
unitary representations of the universal
covering group $\overline{SL(2,\bR)}$. Such representations arise
quite naturally in our consideration, if to realize the Hilbert space
of physical states as an
appropriate function space on the classical phase manifold. We demonstrate
this statement in detail for the case of $C\geq 0$ in subsec. \ref{sQ}.

The paper is organized as follows. Section \ref{s2} provides a
brief description of $\cL$ and $S^1$ as homogeneous spaces of $ISO(1,2)$
in the form most appropriate
for our construction. The consideration follows the general lines used
in  Ref. \cite{KuzenkoLyakhovichSegal95} where a similar description has
been given for the action of $ISO(1,3)$ on sphere $S^2$.

In section 3 we give the Hamilton and Lagrange
formulations for the minimal anyon model with the configuration
manifold $\cM^4=\bR^{1.2}\times S^1$.
The minimal model with the phase space $T^\ast(\bR^{1,2})\times \cL$
is described in section \ref{sL}. For the latter model we show that
in the special case, when the particle momentum is parallel to
the spin vector, the two independent first-class constraints of the theory
split into three ones including second-class constraints.

In section \ref{s5} we discuss the problem of interaction with external
fields in the framework of the minimal anyon models.
We begin with establishing the equivalence between our model
in the special case described and the well known anyon models [11-13,19]
based on the use of the Dirac monopole symplectic two-form.
Thus, along with the latter class of particle models
\cite{ChouNairPolychronakos93,Chou94}, the minimal model in the special case
admits the consistent coupling to an arbitrary week external fields.
In general position, however, the momentum and the spin vector
are not parallel to each other and the minimal model contains essentially two
first-class
constraints. Of course, such a structure of constraints is not compatible
with arbitrary background. Nevertheless,
the model is shown to admit the interaction with
an electromagnetic field subject to the free Maxwell equations with
Chern--Simons term as well as with a constant curvature gravitational
background. The restrictions on the fields appear from the compatibility
conditions for the anyon model gauge symmetries. The similar phenomenon is
known for some superparticle models
\cite{Witten86ShapiroTaylor90Deriglazov93}.


\sect{The Poincar\'e group action on $S^1$ and $\cL$} \label{s2}

In this section we describe the action of $ISO(1,2)$ on
$S^1$ and ${\cL}$ in the form adapted for further consideration.
Let us begin with recalling the realizations of these manifolds as
homogeneous spaces of $SO^\uparrow(1,2)$. It is useful to consider
$S^1$ and $\cL$ as submanifolds of the Riemann sphere
${\bf CP}^1={\bf C}\bigcup\{\infty\}$.
We identify the Lobachevsky plane with the upper half-plane
\begin{equation}
{\cal L}^{(+)}=\{z\in{\bf C},\ \ {\rm Im} \,z>0\} \label{d1}
\end{equation}
or the down half-plane
\begin{equation}
{\cal L}^{(-)}=\{z\in{\bf C}, \ \ {\rm Im} \,z<0\} \label{d2}
\end{equation}
of ${\bf C}$, and realize $S^1$ as $\{z\in{\bf R}\bigcup\{\infty\}\}$,
thus having
${\bf CP}^1={\cal L}^{(+)} \bigcup S^1\bigcup{\cal L}^{(-)}$.
We use both the angle $(\varphi)$ and stereographic
$(z={\bf ctg}\,\frac{\varphi}{2})$ parametrizations of
$S^1$, $\varphi\in\,[0\,,2\pi\,]$.

The Lorentz group $SO^\uparrow(1,2) \cong SL(2,{\bf R})/Z_2$ acts on
${\bf CP}^1$ by fractional linear transformations
\be
N:\ \ z \to z^\prime=\frac{az-b}{d-cz}\ \ ,        \label{d3}
\ee
\begin{displaymath}
N\equiv\left\|N_{\alpha}{}^{\beta}\right\|=
\left (\begin{array}{cc} a & b \\ c & d \end{array}\right)\in SL(2,{\bf R})
\end{displaymath}
With respect to this action ${\bf CP}^1$ consists of three orbits:
${\cal L}^{(+)}$, $S^1$ and ${\cal L}^{(-)}$.

One can bring Eq. (\ref{d3}) to a manifestly covariant form introducing
(by analogy with  the four--dimensional case
\cite{KuzenkoLyakhovichSegal95}) two-component objects
$z^\alpha\equiv(1,z)$ and
$z_\alpha\equiv\epsilon_{\alpha\beta}z^\beta=(-z,1)$
transforming under (\ref{d3}) by the law
\begin{equation}
N:\ z^{\alpha}\to z^{\prime\,\alpha}=\left( \frac{\partial
z^\prime}{\partial z}\right )^{1/2}z^\b N^{-1}{}_\beta {}^\alpha \qquad
\ \bar{z}^\alpha\to \bar{z}^{\prime\,\alpha}=
\left( \frac{\partial \bar{z}^\prime}{\partial \bar{z}}\right )^{1/2}
\bar{z}^\beta N^{-1}{}_\beta {}^\alpha\, , \label{4.a}
\end{equation}
or in the infinitesimal form
\begin{equation}
\delta z=\omega_{\alpha\beta}
z^{\alpha} z^{\beta}\qquad
\delta \bar{z}=\omega_{\alpha\beta}
\bar{z}^{\alpha} \bar{z}^{\beta}\,,\label{4.b}
\end{equation}
where
$\omega_{\alpha\beta}$
are the parameters of Lorentz transformations.

The above relations imply that the Lorentz generators
of scalar fields look like
\begin{equation}
M_a=-i\xi_a\partial_z-i\bar{\xi}_a\partial_{\bar z} \label{d5}
\end{equation}
on ${\cL}$ and
\begin{equation}
M_a=-i\zeta_a\partial_z=-in_a\partial_\varphi \label{d6}
\end{equation}
and on $S^1$ ($z$ being real on $S^1$). Here\footnote{We use Latin
letters to denote vector indices and Greek letters for spinor ones;
the space--time metric signature is chosen to be $(-,+,+)$;
the spinor indices are raised and lowered with the use  of the spinor metric
$\epsilon^{\alpha\beta}=-\epsilon^{\beta\alpha}=
-\epsilon_{\alpha\beta}$ $(\alpha, \beta=0, 1)$, $\epsilon^{01}=1$
by the rule $\psi_{\alpha}=\epsilon_{\alpha\beta}\psi^{\beta}, \psi^{\alpha}=
\epsilon^{\alpha\beta}\psi_{\beta}$.}

\begin{displaymath}
\xi_a\equiv -\frac{1}{2}(\sigma_a )_{\alpha\beta} z^\alpha
              z^{\beta}=-\frac{1}{2}(1+z^2,1-z^2,2z)
\end{displaymath}
\bea
\zeta_a\equiv -\frac{1}{2}(\sigma_a )_{\alpha\beta} z^\alpha
\bar{z}^{\beta}=-\frac{1}{2}(1+z\bar{z},1-z\bar{z},z+\bar{z})
\label{d7}
\eea
\begin{displaymath}
n_a\equiv
(1,-\cos\varphi ,\sin\varphi )\ ,
\end{displaymath}
where
$\partial_z$,
$\partial_{\bar z}$
and $\partial_\varphi$
denote the partial derivatives with respect to $z, \bar{z}$ and $\varphi$.
The three-dimensional Dirac matrices $\sigma_a$ are chosen
to be real and symmetric
\begin{equation}
(\sigma_0)_{\alpha\beta}=
\left( \begin{array}{cc}
		    1 & 0 \\
		    0 & 1
       \end{array}\right)\,\quad
(\sigma_1)_{\alpha\beta}=
\left( \begin{array}{cc}
		    1 & 0 \\
		    0 & -1
       \end{array}\right)\,\quad
(\sigma_2)_{\alpha\beta}=
\left( \begin{array}{cc}
		    0 & 1 \\
		    1 & 0
       \end{array}\right)
\label{d8}
\end{equation}
\begin{displaymath}
(\sigma_a)_{\alpha\beta}(\sigma^a)_{\gamma\delta}=
-(\epsilon_{\alpha\gamma}\epsilon_{\beta\delta}+
\epsilon_{\alpha\delta}\epsilon_{\beta\gamma})\,.
\end{displaymath}

The only Lorentz-invariant objects, constructed in terms of
$z^\alpha$ and $\bar{z}^\alpha$, are the K\"ahler metric on $\cL$ and
the associated two-form
\begin{equation}
{\rm d}s^2=4\frac{{\rm d}z{\rm d}\bar{z}}{\chi^2} \qquad
{\bf\Sigma}=-2i\frac{{\rm d}z\wedge {\rm d}\bar{z}}{\chi^2}\ ,\label{9form}
\end{equation}
where $\chi\equiv\epsilon_{\alpha\beta}z^\alpha\bar{z}^\beta=z-\bar{z}$.
There are no internal Lorentz-invariant structures on $S^1$.
Both $\cL$ and $S^1$ admit external invariants. In particular,
let $p^a$ be a time-like vector. Then the combination
\begin{displaymath}
{\rm d}\sigma\equiv \left|\frac{{\rm d}z}{(p,\zeta)}\right|
\end{displaymath}
remains unchanged under the Lorentz transformations.
For $z=\bar{z}$ we also have
\begin{equation}
{\rm d}\sigma=\frac{{\rm d}\varphi}{(p,n)}  \label{10}
\end{equation}
and ${\rm d}\sigma$ can be treated as a Lorentz-invariant
extension of arc length
(in a rest
reference system where $p^a\sim (1,0,0)$ we have ${\rm d}\sigma\sim
{\rm d}\varphi )$.
As we shall show, this invariant appears in the mechanics Lagrangian
on ${\bf R}^{1,2}\times S^1$  along with the Minkowski
interval $(-dx^a dx_a)^{1/2}$.
Associated with $p^a$ are a number of invariants on $\cL$,
for instance  $(p,\zeta)/\chi$ and ${\rm d}z/(p,\xi)$.

The action of $SO^\uparrow(1,2)$ on ${\bf CP}^1$ can be extended
to that of the Poincar\'e group by defining the translations to act
as the identity map of ${\bf CP}^1$.
As for discrete Lorentz transformations, the mappings on ${\bf CP}^1$
\renewcommand{\theequation}{\thesection.\arabic{equation}a}
\begin{equation}
z\to -\bar{z}\ \ \ \ \ \
\ \ \ \ \ \ \ (z\to-1/\bar{z}) \label{11a}
\end{equation}
can be identified with the space and time inversions of ${\bf R}^{1,2}$
\addtocounter{equation}{-1}
\renewcommand{\theequation}{\thesection.\arabic{equation}b}
\begin{equation}
(x^0,x^1,x^2)\to(x^0,x^1,-x^2)\ \  \  \
(x^0,x^1,x^2)\to(-x^0,x^1,x^2)         \label{11b}
\end{equation}
\renewcommand{\theequation}{\thesection.\arabic{equation}}%
\setcounter{numerone}{\value{equation}}%
respectively. Under the discrete Poincar\'e transformations,
$S^1$ transforms into itself, while ${\cal L}^{(+)}$ turns into
${\cal L}^{(-)}$ and vice versa. Hence ${\bf CP}^1$ consists of two
orbits of the Poincar\'e group, that is,
$S^1$ and ${\cL}={\cL}^{(+)}\bigcup{\cL}^{(-)}$.

\sect{Anyon model on ${\cal M}^4$}
\subsection{Classical dynamics}       \label{s3}

In accordance with the analysis of Sec. 1, the phase space of a minimal
anyon model should be an eight-dimensional transformation space of the
Poincar\'e group. Let us suppose also that the phase space can be
realized as the cotangent
bundle of some manifold $\cM^4$. Then $\cM^4$ proves to have the unique
form ${\bf R}^{1,2}\times S^1$. In this section we construct the particle
dynamics on $\cM^4$.

Let
$A(x^a,p_a,\varphi ,p_\varphi)$ and $B(x^a,p_a,\varphi ,p_\varphi)$
be scalar functions on  $T^\ast(\cM^4)$. In terms of the Poisson bracket
\begin{equation}
\{ A,B\} =\frac{\partial
A}{\partial x^a}\frac{\partial B}{\partial p_a}+ \frac{\partial
A}{\partial\varphi}\frac{\partial B}{\partial p_\varphi}-
(A\leftrightarrow B)\, .
\label{8}
\end{equation}
an infinitesimal Poincar\'e transformation reads
\begin{displaymath}
\delta A=\{ A,-f^a P_a +\omega ^a J_a \}\, ,
\end{displaymath}
where $f^a$ and $\omega^a$ are the parameters of translations
and Lorentz rotations respectively.
The Hamiltonian generators are given by
\begin{equation}
P_a=p_a\qquad J_a=-\epsilon_{abc}x^b p^c +M_a \label{9}
\qquad M_a=n_a p_\varphi\,,
\end{equation}
$\epsilon_{012}=1$, $n_a$ is defined in (\ref{d7}). The particle dynamics is
governed by the first--class constraints
\begin{equation}
P^2+m^2=p^2+m^2\approx 0\, , \label{11}
\end{equation}
\begin{equation}
W=\left(P,M\right)=(p,n)p_\varphi\approx  ms\, , \label{12.a}
\end{equation}
expressing the identical conservation of the mass $m$ and spin $s$
of anyon. For the sake of reparametrization invariance, the
Hamiltonian should be a linear combination of constraints, hence the
action looks like
\begin{equation}
S=\int d\tau\{ p_a
\dot{x}^a+p_\varphi\dot{\varphi}- \frac{e(\tau )}{2}(p^2+m^2)-\lambda
(\tau)((p,n)p_\varphi - ms)\}\, , \label{13}
\end{equation}
where $e(\tau )$ and $\lambda (\tau)$ are Lagrange multipliers, the dots
denote derivatives with respect to evolution parameter $\tau$.
Since the constraint (\ref{12.a}) is linear in the angle momentum
$p_\varphi$, it can be readily solved thus resulting in
the action functional
\begin{equation}
S =\int d\tau\{ p_a \dot{x}^a+\frac{ms}{(p,n)}\dot{\varphi}-
\frac{e(\tau )}{2}(p^2+m^2)\}\, .     \label{14.a}
\end{equation}

The action contains all independent worldline
Poincar\'e invariants on ${\cal M}^4$.
However our manifold also admits the
one-form
\begin{equation}
{\bf\Omega}=\frac{(p,\partial_\varphi n)}{(p,n)}d\varphi \, ,
\label{15}
\end{equation}
that changes under the Poincar\'e transformations only by
exact contributions,
$\delta{\bf\Omega}=d(\partial_\varphi n_a)\omega^a$.
This implies that the one-form
is allowable in the
action functional
\begin{equation}
S=\int d\tau\{p_a\dot{x}^a+\frac{ms}{(p,n)}\dot\varphi
+\varrho\frac{(p,\partial_\varphi n)}{(p,n)}\dot{\varphi}
-\frac{e(\tau)}{2}(p^2+m^2)\} \, ,     \label{14.b}
\end{equation}
where $\varrho$ is a real parameter. This functional
changes by boundary terms under the
Lorentz transformations in contrast to $S$ (\ref{14.a}), the latter being a
genuine invariant. As a consequence, the Noether currents $J_a$
(\ref{9}) take the modified form
\begin{equation}
\tilde{J}_a=-\epsilon_{abc}x^b p^c+\tilde{M}_{a}\,,\ \ \
\tilde{M}_{a}=n_a p_\varphi - \varrho\partial_\varphi n_a\ ,
\label{10.b}
\end{equation}
and the constraint (\ref{12.a}) is modified by the rule
\begin{equation}
P^a \tilde{M}_{a}=(p,n)p_\varphi-\varrho (p,\partial_\varphi n)
\approx ms\ . \label{12.b}
\end{equation}
It is essential that the
constraint (\ref{12.b}) expresses the same physical content as
(\ref{12.a}) has done before, i.e. the strong conservation law of
the anyon spin $s$. This fact is not surprising in so far as
one can show that the models (\ref{14.a}) and (\ref{14.b}) are related
to each other by some canonical transformation with the generating
function
\begin{equation}
{\cal F}(x^a,\varphi ,\tilde{p}_a, \tilde{p}_\varphi )=\varrho
\int\limits_{\varphi_0}^{\varphi}\frac{(\tilde{p},\partial_\varphi n)}
{(\tilde{p},n)}d\varphi+ \varphi\tilde{p}_\varphi+x^a\tilde{p}_a\ ,
\label{generatefunction}
\end{equation}
where the variables without tilde are
related to the model (\ref{14.a}), whereas those with tilde --
to (\ref{14.b}), and $\varphi_0$ is an arbitrary constant.  The
interpretation of the parameter $\varrho$ is also obvious. From
(\ref{10.b}) it follows that the value of $\varrho$ fixes
the value of squared spin $\tilde{M}_{a}\tilde{M}^a=\varrho^2$
(thus the spinning momentum vector is not time-like).

Let us turn to the Lagrangian formulation of the model.
The momenta and the Lagrange multiplier can be eliminated from
(\ref{14.b}) by making use of the mass shell constraint and
the equations of motion. As a result, one  obtains
the following Lagrangian
\begin{equation}
L=-m\sqrt{-\dot{x}^2\left (1-
\frac{2s}{m}\frac{\dot{\varphi}}{(\dot{x},n)}-
\frac{\varrho^2}{m^2}
\frac{\dot{\varphi}^2}{(\dot{x},n)^2}\right )}+
\varrho\frac{(\dot{x},\partial_\varphi n)}{(\dot{x},n)}\dot{\varphi}\ .
\label{16}
\end{equation}
In the limit  $s\to 0, \varrho\to 0$
$L$ reduces to the
Lagrangian of a spinless particle.

The geometric anyon model on $\cM^4$ with $\varrho \neq 0$
can be regarded as a reduction
to $1+2$ dimensions of the $(1+3)$-dimensional spinning particle model
suggested in Ref.
\cite{LyakhovichSegalSharapovUniv}. The latter model contains two parameters
analogous to $\varrho$ and $s$ and possesses an interesting property.
For special relation between $\varrho$ and $s$ the structure
of constraints is radically altered: a first class constraint splits into
pair of second-class constraints (what may be relevant in the framework
the problem of coupling to external fields). In three dimensions, however,
$\cM^4$ does not admit similar phenomenon. Moreover,
for all values of $\varrho$ the corresponding models are
canonically equivalent.

Let us discuss the dynamics in the model.
The equations of motion in Minkowski space can be written
in the form
\be
\dot{p}_a=0 \qquad
\dot{x}^a=e p^a+ms \frac{n^a}{(p,n)^2}\dot{\varphi}-
\varrho \frac{\epsilon^{abc}p_b n_c}{(p,n)^2}\dot{\varphi} \ .
\label{motioneq}
\ee
The rest equations turn out to be identities under the
above equations supplemented by the constraints.
This could be expected, since the anyon possesses as many degrees of freedom
as the spinless particle.

There exist some global restrictions
on the world lines, which are related with causal requirements.
Really, the causal conditions
\begin{equation}
\dot{x}^2<0\ ,\ \ \dot{x}^0>0 \label{causal}
\end{equation}
fulfil on the mass shell if and only if the following inequalities
take place
\renewcommand{\theequation}{\thesection.\arabic{equation}a}
\begin{equation}
s - \sqrt{s^2+\varrho^2}<
\frac{\varrho^2}{m}\frac{\dot{\varphi}}{(\dot{x},n)}<
s + \sqrt{s^2+\varrho^2}\ ,
\end{equation}
for $\varrho\neq 0$ and
\renewcommand{\theequation}{\arabic{equation}}
\addtocounter{equation}{-1}
\renewcommand{\theequation}{\thesection.\arabic{equation}b}
\begin{equation}
\frac{2s}{m}\frac{\dot{\varphi}}{(\dot{x},n)}<1\ ,
\label{causalspecial}
\end{equation}
\renewcommand{\theequation}{\thesection.\arabic{equation}}%
for $\varrho =0$. It is interesting that Lagrangian (\ref{16}) is well defined
only if all the conditions are fulfilled.

Let us describe the general solution of the equations of motion
under the gauge condition $\dot{x}^0=1$. As the Lagrangian
has two independent gauge symmetries, this gauge condition leaves an
arbitrary function $\varphi(t)$ to enter the general solution. The explicit
form of the solution is as follows:
\begin{equation}
\vec{x}(t)=\frac{\vec{p}}{p^0} \left [ {t + ms
\int\limits_{\varphi (0)}^{\varphi (t)} \frac{d\varphi}{(p,n)^2}-
\varrho\int\limits_{\varphi (0)}^{\varphi (t)}d\left (
\frac{1}{pn}\right )} \right ]+ \label{99}
\end{equation}
\begin{displaymath}
+ ms \int\limits_{\varphi (0)}^{\varphi(t)}
\frac{\vec{n}d\varphi}{(p,n)^2}- \varrho\int\limits_{\varphi(0)}^
{\varphi (t)}d\left( \frac{\vec{n}}{p,n}\right )+\vec{x}(0)\ ,
\end{displaymath}
where
\begin{displaymath}
\vec{n}=(-\cos{\varphi}, \sin{\varphi})\qquad
\vec{p}=const\qquad
p^0=\sqrt{\vec{p}^{\ 2}+m^2}\,.
\end{displaymath}
Thus one can see from
(\ref{99}) that, similar to the four-dimensional model
\cite{KuzenkoLyakhovichSegal95} (and also in accordance with the
$3d$-analysis of the paper \cite{CortesPlyushchay95hep}),
the uniform velocity motion of the particle
$\vec{x}(t)=\vec{p}t/p^0$ is superimposed by the
Zitterbewegung with the amplitude of the order of the de
Broglie wave length. This Zitterbewegung can
be eliminated by an appropriate choice of the frame being fixed by
an appropriate gauge condition, e.g.: $\varphi(t)=\varphi(0)$,
unlike the case of $d=1+3$, where the Zitterbewegung is
not a pure gauge artifact \cite{KuzenkoLyakhovichSegal95}.

Now we would like to discuss the relationship of our model with some
known anyon models.
For $\varrho=0$ there exists, in the Hamiltonian formulation,
a local canonical transformation to twistor variables
of the form
\begin{equation}
z\,,\,p_z\to q^\alpha=z^\alpha\sqrt{-2 \varsigma p_z}\ , \qquad
\{q^\alpha,q^\beta\}= \varsigma\epsilon^{\alpha\beta}\, ,  \label{18}
\end{equation}
where $\varsigma = sign\,s$.
As a result, the action (\ref{13}) can be locally transformed into the
twistor action of `semions' \cite{SorokinVolkov94}:
\begin{equation}
S=\int d\tau\left\{p_a \dot{x}^a -
\frac{\varsigma}{2}q_\alpha\dot{q}^\alpha- \frac{e(\tau )}{2}(p^2+m^2)
+ \lambda (\tau)\left(\frac{\varsigma}{4}(p^a\sigma_a)_{\alpha\beta}
q^\alpha q^\beta + ms\right)\right\} \label{19}
\end{equation}
At the same time, the internal phase spaces,
$T^\ast(S^1)$ and the real plane
parametrized by the twistor variables $q^\alpha$, have different
topological structures. Thus the models (\ref{13}) and (\ref{19})
are not globally equivalent. This becomes most transparent at the quantum
level where the spin spectrum turns out to be discrete for
the Sorokin-Volkov model (semions) \cite{SorokinVolkov94}
and continuous for our model.
Let us now consider a global canonical transformation such that the
Pauli-Lubanski function $W$ appears as one of the canonical variables
\be
W =(p,n)p_\varphi - \varrho (p,\partial_\varphi n) \qquad
\Psi = \int\limits_{\varphi\protect_0}^\varphi \frac{d\varphi}{(p,n)}
\label{varactionangle}
\ee
\begin{displaymath}
\tilde{p}_a=p_a\qquad \tilde{x}^a=x^a-W\int\limits^\varphi _{\varphi
\protect _{0}}
\frac{n^a d\varphi}{(p,n)^2}-\varrho\int\limits^\varphi _{\varphi
\protect _{0}}\frac{\epsilon^{abc}p_b n_c}{(p,n)^2} d\varphi \ .
\end{displaymath}
In other words, we describe the spin in terms of action-angle
variables: $W$ is strongly conserved and $\Psi$ is a pure gauge degree
of freedom. Originally these variables were used by
Plyushchay \cite{Plyushchay92} to parametrize the phase space
of the minimal model derived from an extended one by
reducing the dynamics to special constrained surface.
But such a reduction was accompanied by the loss of manifest covariance.
Our consideration shows that the manifest covariance can be restored
by passing to the variables $z, p_z$ we use from the outset.

\subsection{Quantization } \label{sQ}

Let us consider canonical quantization of the model in the framework of the
Dirac method \cite{Dirac64}. This normally implies to perform the
following. All the phase space variables should be defined as
operators subject to the canonical commutation relations in a
Hilbert space, while the physical state subspace is extracted by
imposing the condition that physical wave functions should be annihilated by
the operators of the first-class constraints. It is necessary also to
supply the physical subspace with a well-defined inner product.

In the present model, it is naturally used to realize the Hilbert space of
one-particle states as a space of scalar fields on the
configuration manifold ${\cal M}^4={\bf R}^{1,2}\times S^1$\,.
Then the Poincar\'e group generators are realized in momentum
representation as:
\begin{equation}
{\cal P}_a=p_a\qquad {\cal J}_a=i\epsilon_{abc}p^b
\frac{\partial}{\partial p_c} -in_a\partial _\varphi+s n_a
+\varrho\partial _\varphi n_a   \label{21}
\end{equation}
Now we can realize the first-class constraints
as operators and define the physical states $\Psi (p,\varphi)$ by
the following equations
\begin{eqnarray}
& (p^2+m^2)\Psi (p,\varphi)=0\, \label{20.a} \\ &
((p,\cJ)-ms)\Psi(p,\varphi)=0\, .   \label{20.b}
\end{eqnarray}
For any two physical states $\Psi_1$ and $\Psi_2$ the
Poincar\'e-invariant scalar product is defined by
\begin{equation}
<\Psi_1|\Psi_2>=N\int\limits_{-\infty}^{\infty}\frac{d^2\vec{p}}{p^0}
\int\limits_0^{2\pi} \frac{d\varphi}{(p,n)}\overline{\Psi_1(p,\varphi)}
\Psi_2(p,\varphi),  \label{22}
\end{equation}
with $ p^0=\sqrt{\vec{p}{\ ^2}+m^2}$ and $N$ a normalization constant.

We obtain the well-known realization of the fractional spin
representations \cite{Plyushchay92,CortesPlyushchay95hep}.
The one-particle wave function of
anyon (up to the factor of $(pn)^{1/2}$) is transformed on $S^1$ by
the irreducible unitary representation of the principal continuous
series of group $\overline{SO^\uparrow (1,2)}$
\cite{GelfandGraievVilenkinbook66,Perelomovbook87}. Thereby the choice
of the representation weight should be co-ordinated with the particle
spin. The space of representation is infinite-dimensional and the
constraint equation (\ref{20.b}) has the meaning of the projection onto
the corresponding one-dimensional subspace.
Our approach is also agreement with the results of Ref.
\cite{Plyushchay92}
where the quantization has been fulfilled with the use of
action-angle-type variables (\ref{varactionangle}).

The equations (\ref{20.a}) and (\ref{20.b}) can be explicitly resolved
as follows
\begin{equation}
\Psi (p,\varphi )=C(\vec{p}) exp\left \{-2is\,{\rm arctg}
\left ( {\frac{m\epsilon_{\alpha\beta}z^\alpha z_0^\beta}
{p_{\alpha\beta}z^\alpha z_0^\beta}}\right )-is\varphi
-i\varrho\ln\frac{(p,n)}{(p,n_0)}\right \}
\label{23}
\end{equation}
where
\begin{displaymath}
p^0=\sqrt{\vec{p}{\ ^2}+m^2}\qquad
p_{\alpha\beta}\equiv(p^a\sigma_a)_{\alpha\beta}
\qquad z^\alpha=(1,{\rm ctg}\frac{\varphi}{2})\ ,\
\end{displaymath}
and $\varphi_0$ is the integration constant of Eq. (\ref{20.b}).

Let us point out that the generators $\cJ_a$ in (\ref{21}) are chosen so
that the wave functions (\ref{23}) are well-defined on $\cM^4$. There
are possible different realizations for $J_a$. In particular,
the third term in the expression (\ref{21}) for $J_a$ can be removed
by an unitary transformation
\cite{CortesPlyushchay95hep}. However the resulting
wave function
\be
\tilde\Psi=e^{is}\Psi
\ee
turns out to be multivalued on $S^1$.

It is worth discussing the relationship between the quantization of
the minimal model for $\varrho=0$ and
the model of `semions' \cite{SorokinVolkov94} described by the
action (\ref{19}). These mechanical models are related
at the classical level through
the local canonical transformation (\ref{18}), but their
phase spaces have different topology.
That is why the models possess different spin spectra.
The quantization of the model
(\ref{19}) has been realized in Ref. \cite{SorokinVolkov94} on the
bounded below representations of the discrete series with lowest
weights $1/4$ and $3/4$.
Only the particles with
spin $s=(2n+1)/4$, $n$ being a positive integer,
(so called semions) are allowed to appear in spectrum of model
\cite{SorokinVolkov94}.

Finally, we consider in more detail the description of the case of
(half) integer spin in the framework of our geometric construction. In
that case the wave function of anyon (\ref{23}) can be expanded on
$S^1$ in relativistic harmonics of the form $z^\alpha /(p\xi)^{1/2}$.
For spin $s =\pm k/2\ (k=0,1,2 \ldots )$ one gets
\begin{displaymath}
\tilde\Psi=exp(-i\varrho\ln(pn))
F_{\alpha_1\alpha_2\ldots\alpha_k}(p)\frac{z^{\alpha_1}z^{\alpha_2}\ldots
z^{\alpha_k}}{(p,\xi)^{k/2}}\ ,
\end{displaymath}
and Eqs. (\ref{20.a},\ref{20.b}) are equivalent
to the Dirac equation for
$F_{\alpha_1\alpha_2\ldots\alpha_k}(p)$
\begin{equation}
(p_{\alpha_1}{}^{\beta}+im\varsigma\,
\delta_{\alpha_1}{}^{\beta}) F_{\beta\alpha_2\ldots\alpha_k}(p)=0
\end{equation}
\begin{displaymath}
F_{\alpha_1\alpha_2\ldots\alpha_k}=F_{(\alpha_1\alpha_2\ldots\alpha_k)}\qquad
\vert s\vert =\frac{k}{2}\qquad \varsigma=sign\,s\, .
\end{displaymath}
Thus, the (half) integer spin particle admits the description
in terms of usual finite-component fields transforming by
the finite-dimensional representation of $SL(2,\bR)$.

\sect{The anyon model on $T^\ast({\bf R}^{1,2})\times{\cal L}$}\label{sL}

We turn now to the case $C<0$. Then the spin space has to be
homeomorphic to Lobachevsky plane. So the anyon dynamics can be realized
on the phase space $T^\ast({\bf R}^{1,2})\times{\cal L}$.

Let us start with the following action functional
\begin{equation}
S=\int(p_adx^a+\varrho{\bf\Omega}(z,\bar{z})-H(x^a,p_a,z,\bar{z},e,\lambda)d\tau)
\label{L1}
\end{equation}
\begin{displaymath}
{\bf\Omega}(z,\bar{z})=\frac{i}{\chi}(dz+d\bar{z})\,,
\end{displaymath}
where the Hamiltonian
\begin{displaymath}
H=e(\tau)T_1+\lambda(\tau) T_2
\end{displaymath}
presents a linear combination of the constraints
\begin{equation}
T_1=p^2+m^2\approx 0 \qquad
T_2=2i\varrho\frac{(p,\zeta)}{\chi}-ms\approx 0 \,.
\label{L2}
\end{equation}
Here $e(\t)$ and $\l(\t)$ are Lagrange multipliers. It follows from
(\ref{L1}) that the symplectic structure on the manifold is determinated
by the Poincar\'e-invariant two-form
\be
{\rm d}p_a\wedge{\rm d}x^a+\varrho{\bf\Sigma}  \label{L3}
\ee
with ${\bf\Sigma}={\rm d}{\bf\Omega}$ a Poincar\'e-invariant two-form on the
Lobachevsky plane. Due to (\ref{L3}), we are able to identify the phase space
with $T^\ast({\bf R}^{1,2})\times{\cal L}$. The non-vanishing Poisson
brackets read
\begin{equation}
\{x^a\,,p_b\}=\d^a{}_b \qquad \{z\,,\bar{z}\}=\frac{i}{2\varrho}(z-\bar{z})^2\,.
\label{L5}
\end{equation}
It should be noted that the action functional ({\ref{L1}) is not
Poincar\'e-invariant since the one-form $\Omega$ changes on an
exact one-form under an infinitesimal Lorentz transformation
\begin{equation}
\d_{\omega}\Omega=\frac{i}{2}\omega^a{\rm d}(\partial\xi_a-\bar{\partial}
\bar{\xi}_a)\,. \label{L6}
\end{equation}
To reveal the physical content of the model, we consider the Hamilton
generators of Poincar\'e transformations
\begin{equation}
{\cal P}_a=p_a\qquad {\cal J}_a=\epsilon_{abc}x^b p^c+ M_a\;,\label{L7a}
\end{equation}
\begin{equation}
M_a=2i\varrho\frac{\zeta_a}{\chi}=
-i\varrho\left(\frac{1+z\bar{z}}{z-\bar{z}}\;,\;
\frac{1-z\bar{z}}{z-\bar{z}}\;,\;\frac{z+\bar{z}}{z-\bar{z}}\right)\,.
\label{L7}
\end{equation}
Comparing the latter and (\ref{L2}) we show
that the model describes an irreducible dynamic of the particle
with mass $m$, spin $s$ and timelike spin vector,
$M_a M^a=-\varrho^2$.

The model on $T^\ast({\bf R}^{1,2})\times{\cal L}$ implies some restrictions
on the parameters $\varrho$ and $s$. Using the identity
\begin{equation}
u_a=4\frac{(u,\zeta)}{\chi^2}\zeta_a
-2\frac{(u,\xi)}{\chi^2}\bar{\xi_a}
-2\frac{(u,\bar{\xi})}{\chi^2}\xi_a
\end{equation}
for arbitrary 3-vector $u_a$, we get
\begin{equation}
u^2=
4\frac{(u,\zeta)^2}{\chi^2}+4\left|\frac{(u,\xi)}{\chi}\right|^2
\label{ident}
\end{equation}
Now, applying the last identity for $\m^a=p^a$ and accounting the
constraints (\ref{L2}), we arrive at
\begin{equation}
s^2-\varrho^2=\left|\frac{2\varrho}{m}\frac{(p,\xi)}{\chi}\right|^2
\label{L8}
\end{equation}
Thus the constraints (\ref{L2}) are non-contradictory only under the
restriction $k=\left|s/\varrho\right|\geq 1$.
Moreover, for $k=1$ the momentum and the spin vector become parallel
to each other
\begin{equation}
p_a=2im\frac{\zeta_a}{\chi}\,.  \label{L9}
\end{equation}
As a consequence, for $\varrho=\pm s$  the structure of the
constraints is drastically changed: instead of two first-class
constraints we get two second-class constraints and one first-class
constraint. In the following section we shall show that for $|k|=1$
the model can be treated as a version of the well-known anyon model
with Dirac's monopole two-form [11-14,19]

By eliminating $p^a$ and the Lagrange multipliers from (\ref{L1}), we
get the Lagrangian
\begin{equation}
L=-2m\sqrt{k^2-1}\left|\frac{(\dot{x},\xi)}{\chi}\right|
-2imk\frac{(\dot{x},\zeta)}{\chi}+i\varrho\frac{\dot{z}+\dot{\bar{z}}}{\chi}\ ,\
k=\frac{s}{\varrho}\,, \label{L10}
\end{equation}
where the identity (\ref{ident}) has been used. From here we again get the
restriction $|k| \geq 1$. For $|k| > 1$ the action possesses two gauge
symmetries
\bea
\delta_{\epsilon_1}x^a&=&2p^a\epsilon_1 \qquad \;\;\;\;
\delta_{\epsilon_1}z=
\delta_{\epsilon_1}\bar{z}=0\,; \\ \nonumber
\delta_{\epsilon_2}x^a&=&2i\varrho
\frac{\zeta^a}{\chi}\epsilon_2 \qquad
\delta_{\epsilon_2}z=-i(p,\xi)\epsilon_2 \qquad
\delta_{\epsilon_2}\bar{z}=i(p,\bar{\xi})\epsilon_2\,.
\label{L11}
\eea
Here $p^a$ is the three-momentum
\begin{displaymath}
p_a\equiv\frac{\partial L}{\partial\dot{x}^a}=-m\frac{\sqrt{k^2-1}}{|\chi|}
\left[\sqrt{\frac{(\dot{x},\bar{\xi})}{(\dot{x},\xi)}}\ \xi_a+
\sqrt{\frac{(\dot{x},\xi)}{(\dot{x},\bar{\xi})}}\ \bar{\xi}_a\right]
-2imk\frac{\zeta_a}{\chi}\,.
\end{displaymath}
For $|k|=1$, when
$p_a=2im\zeta_a/\chi\, ,\,(p,\xi)=(p,\bar{\xi})=0$,
the symmetries become dependent. This is a consequence of the fact
that for $|k|=1$ there remains only one first-class constraint.

Let us give a few comments on the dynamics. Similarly to the model
on $\cM^4$, here in general position $|k| > 1$ we have the Zitterbewegung and
generalized causality conditions. The case $|k|=1$ is again very special.
Here and only here the Zitterbewegung is absent: the particle moves
along a straight line in ${\bf R}^{1,2}$ and remains in rest on $\cL$,
the position on $\cL$ being determined by the constraints.

\sect{Coupling to external fields}\label{s5}

It is well known that for constrained dynamical systems the
existence of consistent coupling to external fields is not obvious and
deserves special study.
The behavior of the $3d$-spinning particle in arbitrary external
electromagnetic and gravitational fields has been recently
studied in the framework of the model with the Dirac monopole
symplectic two-form \cite{ChouNairPolychronakos93,Chou94}.
In this approach the dynamics
of a particle with mass $m$ and spin $s$ is realized in a six-dimensional
phase space with the symplectic structure
\be
{\rm d}p_a\wedge{\rm d}x^a+\frac{s}{2}\,
\frac{\epsilon^{abc} p_a{\rm d}p_b\wedge{\rm d}p_c}{(-p^2)^{3/2}} \label{M1}
\ee
such that the mass-shell equation (\ref{11})
presents itself the only constraint in the theory. As a consequence,
the model admits coupling to an arbitrary background. From (\ref{M1})
one deduces the fundamental Poisson brackets
\be
\{x^a,x^b\}=-s\frac{\epsilon^{abc}p_c}{(-p^2)^{3/2}}\qquad
\{x^a,p_b\}=\delta^a{}_b\qquad            \label{M2}
\{p^a,p^b\}=0.
\ee

The peculiar feature of the model is that the momentum and the
spin vector are parallel. This is the same condition that is characteristic
of the minimal anyon model for $|k|=1$, when the model contains three
constraints instead of two ones in the general position.
For $|k|=1$ and only in this case the model possesses one first-class
and two second-class constraints, instead of two first-class constraints
in general position. Obviously, only the former constraint's structure
is stable with respect to deformations by  weak background fields.
In fact, for $|k|=1$ the minimal model can be treated as a
reformulation of the monopole model (\ref{M1},\ref{M2})
in special extended phase space.
This can be seen by reducing the dynamics on the surface of the
second-class constraints.

In general position, the model seems to possess a limited number of
admissible backgrounds. We present below two special cases
of external fields.
First, let us consider a minimal coupling of the model to a
curved gravitational background. Then, the constraints turns into
\begin{equation}
T_1=g^{mn}\nabla _m\nabla _n+m^2\approx 0\qquad T_2=
{e_a}^m \nabla _m
M^a-ms\approx 0\,, \label{v1}
\end{equation}
where
\begin{displaymath}
\nabla_m=p_m+\omega_{m\,ab}\epsilon^{abc}M_c\ ,
\end{displaymath}
$\omega_{m\,ab}$ being the torsion-free spin connection
and ${e_m}^a$ a dreibein associated to the metric $g_{mn}$.
However, not every background field preserves the structure
of the constraints. Using the identities
\begin{displaymath}
\{\nabla_a,\nabla_b\}=(\eta_{ab}R_{bd}+\eta_{bd}R_{ac}-\frac{1}{2}R\eta_{ac}
\eta_{bd})\,\epsilon^{cde}M_e\ ,
\end{displaymath}
where $R_{ab}={e_a}^m {e_b}^n R_{mn}$, $R=R_a{}^a$ are Ricci tensor
and scalar curvature respectively, one can arrive at the following
commutation relation
\be
\{T_1,T_2\}=-2\epsilon^{abc}M_a R_{bd} M^d {e_c}^m \nabla_m\label{v2}.
\ee
This expression vanishes if and only if the space-time has a constant
curvature, that is $R_{mn}=\frac{1}{3}Rg_{mn}$.
The constraints (\ref{v1}) could also be obtained as a
dimensional reduction of the $d=1+3$ anti de Sitter spinning
particle model \cite{KuzenkoLyakhovichSegalSharapovAdS}.

Next, we consider the interaction of anyon with an external $U(1)$
-\,gauge field, using the following covariant generalization of momentum
\cite{LatinskySorokin91}:
\begin{displaymath}
p_a\to P_a=p_a+eA_a(x)+lF_a(x)\,,
\end{displaymath}
where $A_a(x)$ is the Abelian gauge potential,
$F_a(x)=1/2\epsilon_{abc}F^{bc}=\epsilon_{abc}\partial ^bA^c$
is the dual field strength, $e$ and $l$ are the charge and anomalous
magnetic momentum of particle respectively. Then the Poisson bracket
of the constraints
\begin{displaymath}
T_1=P^2+m^2\approx0 \qquad T_2=(P,M)-ms\approx 0
\end{displaymath}
transforms to
\be
\{T_1,T_2\}=2P_aM_b\epsilon^{abc}(eF_c+l\partial^d F_{cd})\,.
\label{Chern-Simons}
\ee
It is seen that the
bracket of constraints is proportional to the free equations of
Chern-Simons topological massive electrodynamics, with
$e/l$ the mass of the field.
The r.h.s. of (\ref{Chern-Simons}) should
vanish to preserve the constraints structure.  In
this way the free equations of the Chern-Simons
electrodynamics emerge from the anyon model as compatibility
conditions of the particle constrained dynamics. The topological mass
of the field appears to be consistent with charge and anomalous
magnetic momentum ratio for the anyon.  The similar phenomenon is
recognized in some superparticle and superstring models, where the
massless field equations are reproduced for the {\it external fields}
from the consistency requirements of the model dynamics
\cite{Witten86ShapiroTaylor90Deriglazov93}.

In the case when the particle spin is not parallel to the momentum,
one can see from (\ref{v2},\ref{Chern-Simons})
that the homogeneous gravity background and the
free Chern-Simons Abelian gauge field probably exhaust all the backgrounds,
allowing the consistent coupling to the anyon.

\sect{Concluding remarks} \label{s6}

In this paper the minimal model of anyon has been considered in depth.
Although the action of the $d=3$ Poincar\'e group is nonlinearly realized
on the spin phase space ($T^\ast(S^1)$ or $\cal L$),
the suggested geometric construction
provides the manifestly covariant formulation without any auxiliary
variables introduced. Besides the formulation has the
transparent geometric sense, it can give an efficient tool for the study of
the quantization and coupling of the anyon to external fields.
It should be noted that these problems are allowed to be treated
by means of this construction without employing the known methods
such as induced representations and Dirac-monopole two-form [10-14],
which can not be applied in general case, when the
particle momentum is not parallel to the spin vector.

We restricted our consideration of the quantum theory to the case
of the phase manifold $T^\ast({\bR}^{1,2}\times S^1)$.
Here the quantization of the minimal model naturally  leads
to the well known realization of quantum anyon theory on the fields
transforming by the unitary representations of the continuous series of
$\overline{SO(1,2)}$. The quantization on
the manifold $T^\ast({\bR}^{1,2})\times \cL)$ could be performed in
a similar way. The only essential distinction is that the realization of
the unitary representation of the Poincare-group requires to use
the appropriate infinite-dimensional representation of the {\it
discrete} series of Lorentz group. It is remarkable that the spinning
part (\ref{L7}) of the Lorentz generators (\ref{L7a}) coincides with
the Berezin's symbols of generators of the corresponding representations
used by geometric quantization on $\cL$ \cite{Berezin75,Perelomovbook87}.
Thus, for the quantization in the spin subspace one can use
the well-studied method of the geometric quantization.

\vspace*{1cm}
\noindent
{\large\bf Acknowledgments}
Two of us (IVG and SLL) are thankful to A.Yu. Segal, A.A. Sharapov
and K.M. Schechter for helpful discussions. This work was supported
in part by European Community grant INTAS No 93-2058.

\end{document}